# From Data-Driven Models to Physical Insight: Vibrational Entropy Governed by Atomic Volume

Shivam Tripathi*, Jatin Kawatra, Varun Malviya, Krishna Mehta

Department of Materials Science and Engineering, Indian Institute of Technology Kanpur, Kanpur, Uttar Pradesh, India 208016

*Corresponding author: Shivam Tripathi (shivamt@iitk.ac.in)

# Abstract

Vibrational entropy plays a central role in determining phase stability and temperature-dependent behavior in materials, yet its calculation from first-principles phonon methods remains computationally demanding. In this work, we combine data-driven modeling with physically motivated analysis to develop an efficient and interpretable framework for predicting vibrational entropy. Using a dataset derived from PhononDB, a feedforward neural network trained on Materials Project and composition-based descriptors achieves high predictive accuracy, while SHAP analysis identifies atomic volume as the dominant factor governing vibrational entropy. Guided by this insight, simplified analytical models are constructed, revealing a logarithmic dependence of vibrational entropy on atomic volume consistent with lattice dynamical considerations. A logarithmic–linear model is shown to provide an accurate and physically interpretable description across the full range of materials.

To extend the analysis to finite temperatures, a temperature-dependent formulation is introduced that incorporates $T^3$ scaling at low temperatures and logarithmic dependence at higher temperatures, consistent with Debye- and Einstein-type behavior. This unified model captures both structural and thermal contributions to vibrational entropy with good accuracy. Overall, the proposed framework demonstrates that vibrational entropy can be predicted using simple, physically meaningful relationships, offering a computationally efficient alternative to full phonon calculations and enabling entropy-informed materials screening

# 1. Introduction

Vibrational entropy ($S_{vib}$) is a fundamental thermodynamic quantity that plays an important role in determining the stability and temperature-dependent behavior of materials. As a key component of the free energy, it directly influences phase equilibria, solid–solid transformations, defect formation, and diffusion processes. In many systems, particularly at elevated temperatures, vibrational entropy contributions can rival or even dominate enthalpic effects, thereby governing phase selection and stability [1–6]. This is especially relevant in complex materials such as high-entropy alloys [7,8], thermoelectric materials [9-11], and metastable phases [12-15], where entropy-driven stabilization has emerged as a central design principle.

At the microscopic level, vibrational entropy originates from lattice dynamics and is determined by the phonon density of states [3-5]. The distribution of phonon modes depends sensitively on atomic arrangement, bonding characteristics, and mass distribution [4,5]. Consequently, even subtle variations in structure or composition can lead to measurable differences in $S_{vib}$[16-18]. Accurately capturing these effects is essential for predictive thermodynamics and for understanding the finite-temperature behaviour of materials.

First-principles methods based on density functional theory (DFT), combined with lattice dynamics calculations, provide a rigorous framework for computing vibrational entropy. Two primary approaches are commonly employed: finite-displacement (supercell) method [19] and density functional perturbation theory (DFPT) [20]. In the finite-displacement approach, large supercells are constructed and multiple atomic displacements are introduced sequentially to compute second-order interatomic force constants. In contrast, DFPT evaluates these force constants through linear-response calculations, but still requires solving perturbations for each atomic degree of freedom and wave vector. In both cases, the workflow is inherently sequential and computationally intensive: force constants must first be obtained, followed by phonon dispersion calculations over dense Brillouin-zone meshes, and finally integration to obtain thermodynamic quantities such as $S_{vib}$. While quite accurate, these approaches scale steeply with system size, symmetry reduction, and chemical complexity. Low-symmetry or multicomponent materials further increase the number of required perturbations or displacements, significantly amplifying computational cost. Consequently, for large-scale materials discovery efforts involving thousands of compounds, such phonon calculations become prohibitively expensive in terms of both computational time and resource allocation, creating a major bottleneck for incorporating $S_{vib}$ into high-throughput workflows.

Recent advances in data-driven approaches offer a promising route to address this challenge, enabling rapid prediction of vibrational properties from material descriptors. Graph neural network–based models such as atomistic line graph neural network (ALIGNN) [21] and Euclidian neural network [22] architectures have demonstrated high accuracy in predicting phonon density of states and derived quantities, including $S_{vib}$, directly from crystal structures. However, these approaches require detailed structural information, including atomic coordinates and bonding environments, and rely on large training datasets generated from computationally expensive phonon calculations, making them challenging to apply in systems with uncertain or disordered structures, such as doped or experimentally realized materials. In contrast, composition-based models have shown that $S_{vib}$ can be predicted using only elemental information, even with relatively small datasets (~300 samples) [23], but such models inherently lack structural resolution and cannot distinguish between different polymorphs or phases of the same composition. Consequently, despite their predictive capability, existing approaches either sacrifice interpretability and practicality or fail to capture essential structural effects, highlighting the need for models that balance accuracy, simplicity, and physical insight.

In this work, we take an approach that combines first-principles data, machine learning, and simple physics-based modeling to predict $S_{vib}$ in an efficient and interpretable way. Starting from a large dataset derived from PhononDB [24,25], we build a feature set using Materials Project descriptors [26] and Magpie features [27], and train a feed forward neural network to learn the underlying trends. Using shapely additive explanations (SHAP) analysis [28], we find that atomic volume plays a dominant role in governing $S_{vib}$. Guided by this observation, we explore simplified analytical models, including linear, logarithmic, and logarithmic–linear forms, to better understand the relationship between vibrational entropy and atomic volume. While the linear and logarithmic models capture limiting trends in different regimes, the logarithmic–linear form provides a more consistent description across the full range of atomic volumes by combining the dominant logarithmic dependence with a linear correction. This hybrid representation retains physical interpretability while improving accuracy, capturing both the primary trend and systematic deviations arising from variations in material properties. This analysis is further extended to incorporate temperature dependence, where distinct scaling behaviors are expected from lattice dynamics. In particular, $S_{vib}$ follows a $T^3$ dependence at low temperatures and transitions to a weaker, approximately logarithmic dependence at higher temperatures. Building on these considerations, a temperature-dependent formulation is developed to capture both regimes within a unified and physically interpretable framework.

Overall, this framework shows that $S_{vib}$, despite its complexity, can be described using simple and physically meaningful relationships. This provides a fast alternative to full phonon calculations and enables more practical, entropy-aware materials screening.

## 2. Methods

### 2.1. Vibrational entropy from phonon calculations

Vibrational entropy was evaluated within the harmonic approximation using second-order interatomic force constants available in the PhononDB dataset [24]. For each material, these force constants are obtained from finite-displacement calculations, where a set of atomic displacements and the corresponding forces are provided through the phonopy.yaml and FORCE_SETS files. From the force constants, phonon frequencies are calculated throughout the Brillouin zone, and the phonon density of states is constructed by sampling a uniform $20 \times 20 \times 20$ q-point mesh. Using phonopy package [19]. This provides a sufficiently dense representation of vibrational modes for evaluating thermodynamic properties.

Within the harmonic approximation, the vibrational entropy at temperature $T$ is given by:

$$S_{vib}(T) = \int_0^\infty g(\omega) \left[ \frac{\hbar\omega}{k_B T} \left( e^{\frac{\hbar\omega}{k_B T}} - 1 \right)^{-1} - \ln\left( 1 - e^{-\frac{\hbar\omega}{k_B T}} \right) \right] d\omega \qquad (1)$$

where $g(\omega)$ is phonon density of states. In this work, the entropy is expressed in units of $k_B$ per atom by normalizing with respect to the number of atoms in the unit cell. All values are reported at T = 300 K unless otherwise specified; temperature-dependent $S_{vib}$ was also evaluated over the range 100-500 K at intervals of 50 K, with finer temperature resolution readily accessible at negligible additional computational cost.

## 2.2. Feature construction

To enable predictive modeling, a diverse set of descriptors was constructed by combining data from the Materials Project [26] and composition-based features generated using Magpie [27]. Materials Project descriptors include structural, thermodynamic, elastic, electronic, magnetic, dielectric, and surface properties. These features provide physically meaningful information related to bonding, stability, and electronic structure, some of which influence lattice dynamics and phonon behavior. In parallel, Magpie descriptors encode compositional information through statistical measures (minimum, maximum, mean, range, average deviation, and mode) of elemental properties such as atomic number, Mendeleev number, electronegativity, valence electron configuration, and melting temperature.

A systematic data cleaning protocol was applied to ensure robustness and eliminate redundancy. First, low-variance features (≥90% identical values) were removed. Second, features with ≥10% missing values were discarded to retain reliable descriptors. Third, any remaining rows containing missing values were removed, resulting in a fully dense dataset. Finally, multicollinearity was reduced using correlation-based filtering: for feature pairs with $|r| \geq 0.8$, pairs were identified from the upper triangle of the correlation matrix to avoid duplicate comparisons, and the feature with the lower absolute correlation with $S_{vib}$ was removed.

Non-numeric features were treated conservatively. Boolean descriptors (e.g., stability indicators) were retained and converted into binary form (0/1), while all other categorical or complex object-type features were excluded to ensure compatibility with machine learning algorithms. After preprocessing, the final dataset consists of 9,866 materials with 59 input features, forming a fully numerical and non-redundant feature space suitable for modeling $S_{vib}$. The complete list of retained features is provided in the Supporting Information.

## 2.3. Neural network model and feature importance analysis

A feedforward neural network was developed to predict $S_{vib}$ from the constructed feature set. The dataset was split into training, validation, and test sets using a 70:10:20 ratio. Specifically, 20% of the data was held out as an independent test set, while the remaining 80% was further partitioned to obtain a validation set corresponding to 10% of the total dataset. The model consisted of a single hidden layer, with the number of neurons varied (2, 4, 8, and 16) to examine the trade-off between model complexity and predictive performance. A rectified linear unit (ReLU) activation function [29] was used in the hidden layer, and a linear activation was used at the output for regression. L1 regularization (coefficient = 0.01) [30] was applied to mitigate overfitting. Training was performed using the Adam optimizer [31] with mean squared error as the loss function, for 100 epochs and a batch size of 16.

Model performance was assessed using mean squared error (MSE), mean absolute error (MAE), and the coefficient of determination ($R^2$) across training, validation, and test sets. Among the architectures considered, the model with two neurons in the hidden layer provided the best balance between accuracy and simplicity. Increasing the number of neurons did not yield a meaningful improvement.

The selected architecture was retrained with early stopping based on validation loss (patience = 20 epochs), allowing training for up to 1000 epochs. In practice, convergence occurred earlier, and the model corresponding to the lowest validation error was retained. Final evaluation on the test set confirmed stable generalization. Training and validation loss curves were examined to ensure consistent convergence, and parity plots were used to assess agreement between predicted and actual values.

To interpret the model, SHAP analysis [28] was performed using the trained two-neuron network. The same data split and feature scaling used during training were retained. The training dataset was used as the background distribution, and SHAP values were computed for the test set. Feature importance was quantified using the mean absolute SHAP value, allowing identification of the most influential descriptors governing vibrational entropy. This analysis provides insight into the physical factors captured by the model and links the learned relationships to meaningful materials descriptors.

## 2.4. Reduced-order modelling using the dominant descriptor

Following the SHAP analysis, 'density_atomic' was identified as the most influential feature governing $S_{vib}$. In the Materials Project database, this quantity corresponds to the atomic volume (Å³/atom) and is therefore used here as a descriptor of local structural packing. To examine how effectively this dominant descriptor alone captures the variation in $S_{vib}$, simplified analytical models were constructed using atomic volume as the sole input. Linear ($S_{vib} = a * x + b$), logarithmic ($S_{vib} = a * ln(x) + b$), and log–linear forms ($S_{vib} = a * ln(x) + b * x + c$) were considered to represent possible functional relationships between atomic volume and $S_{vib}$.

The parameters for all models were obtained using least-squares regression on the training data, following a 70–30 train–test split consistent with the previous models. The performance of models were evaluated using standard regression metrics and parity analysis. In addition, to examine model performance across different atomic volume regimes, the dataset was partitioned into bins based on atomic volume, and the MSE was evaluated within each bin. This enables assessment of how model accuracy varies across the range of atomic volumes.

This approach provides a simplified and physically interpretable framework for describing the dependence of $S_{vib}$ on atomic volume, while enabling direct comparison between different functional forms and with the neural network model.

## 2.5. Extension to temperature-dependent modelling

To capture the temperature dependence of $S_{vib}$, a dataset combining atomic volume, temperature, and $S_{vib}$ was constructed and divided into training and test sets using a 70:30 split. A piecewise modelling approach was adopted to account for distinct low- and high-temperature behaviour. At lower temperatures, vibrational entropy was modelled using:

$$S_{vib} = a_1 * ln(x) + b_1 * x + c_1 * T^3 + d_1 \qquad (2)$$

while at higher temperatures the form:

$$S_{vib} = a_2 * ln(x) + b_2 * x + c_2 * ln(T) + d_2 \qquad (3)$$

was used, where $x$ denotes atomic volume and T represent temperature. $a_i$, $b_i$, $c_i$, and $d_i$ are fitting constants. The transition temperature ($T_c$) separating the two regimes was determined by selecting the value that minimized the training MSE. The corresponding models were then trained within their respective temperature ranges and combined to generate predictions. Model performance was evaluated using MSE, MAE and $R^2$ on both training and test sets, and parity plots were used to assess predictive agreement.

This formulation provides a compact and physically motivated description of temperature-dependent $S_{vib}$, capturing both low-temperature phonon behavior and high-temperature trends within a unified framework.

## 3. Results and Discussion

### 3.1. Neural network prediction and interpretation vibrational entropy

The predictive capability of the neural network model was first evaluated using the full feature set comprising 9866 materials and 59 input descriptors (see SI Table 1). Model architectures with varying complexity, consisting of 2, 4, 8, and 16 neurons in a single hidden layer, were examined to assess the balance between model capacity and generalization. The evolution of the MSE during training is shown in SI Figure 2, and the corresponding performance metrics (MSE, MAE, and R²) are summarized in SI Table 2.

All models exhibited strong predictive performance, with low errors and close agreement between predicted and computed $S_{vib}$. Increasing the number of neurons beyond two resulted in only marginal improvements, indicating that the underlying relationship can be captured using a relatively simple architecture. The model with two neurons was therefore selected and retrained, see Section 2.3, providing a favorable balance between accuracy and model simplicity.

The training and validation loss curves in SI Figure 2 indicate stable convergence without evidence of significant overfitting, with early stopping triggered at epoch 119. Parity plots (Figure 1) further confirm that the model accurately reproduces $S_{vib}$ across the dataset, achieving test-set values of MSE = 0.064 $(k_B/atom)^2$, MAE = 0.180 $k_B/atom$, and $R^2$ = 0.968. These results demonstrate that the selected neural network provides a reliable representation of the relationship between the input descriptors and $S_{vib}$.

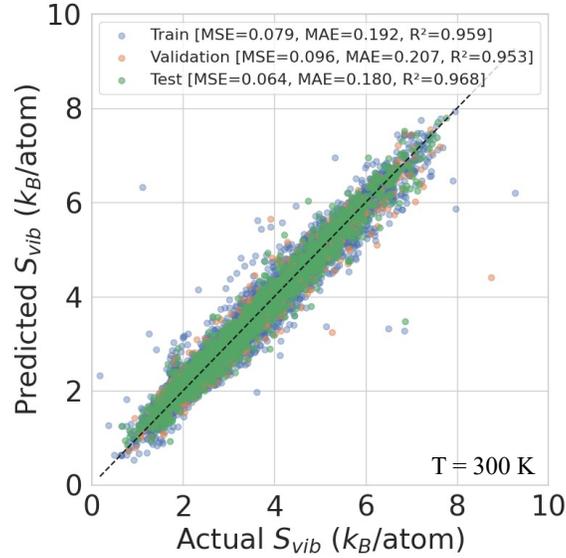

**Figure 1:** Parity plot comparing predicted and actual vibrational entropy ($S_{vib}$) values for the training, validation, and test sets. The mean squared error (MSE) and mean absolute error (MAE) are reported in units of $(k_B/atom)^2$ and $k_B/atom$, respectively.

SHAP analysis, see Figure 2, identifies atomic volume and mean atomic number as the dominant contributors to the predicted $S_{vib}$, with additional influence from formation energy and electronic structure descriptors. This highlights the primary role of atomic packing and elemental characteristics in governing lattice dynamics.

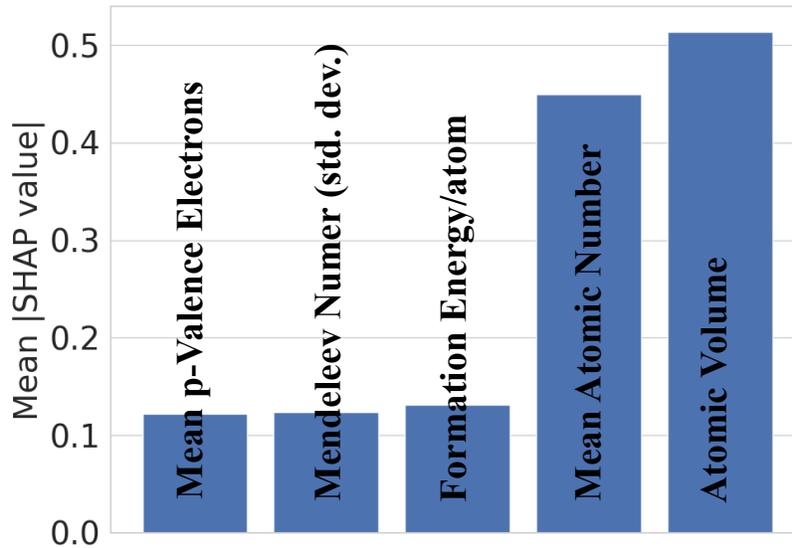

**Figure 2:** Top five features ranked by mean absolute SHAP values, demonstrating that atomic volume and mean atomic number are the primary contributors to the predicted vibrational entropy, followed by formation energy and electronic structure descriptors.

## 3.2. Dependence of vibrational entropy on atomic volume

The relationship between $S_{vib}$ and atomic volume was examined using linear, logarithmic, and logarithmic–linear models, as shown in Figure 3(a). All three models capture the overall increasing trend of $S_{vib}$ with atomic volume, but clear differences in their behavior are observed across the range. At lower atomic volumes, the logarithmic model follows the data more closely than the linear model, capturing the pronounced nonlinear increase in $S_{vib}$. The logarithmic–linear model exhibits similar behavior in this regime and provides a comparable level of agreement. In contrast, the linear model deviates due to its inability to represent the curvature present at small atomic volumes. At larger atomic volumes, the trend approaches a near-linear dependence. In this regime, the linear model provides a reasonable approximation, while the logarithmic model tends to underestimate the increase. The logarithmic–linear model again remains consistent with the data, closely tracking the linear trend without the underestimation observed in the logarithmic model. The logarithmic–linear model therefore provides a continuous description across the entire range, effectively bridging the nonlinear behavior at low atomic volumes and the near-linear scaling at higher values, without exhibiting the systematic deviations observed in the individual linear and logarithmic models.

To further quantify model performance across different regimes, the MSE was evaluated within atomic volume bins, as shown in Figure 3(b). At low atomic volumes, the logarithmic model shows lower error than the linear model, while the logarithmic–linear model achieves the lowest or comparable MSE in this region. At intermediate volumes, all three models exhibit similar performance, with the logarithmic–linear model consistently remaining among the lowest in error. At higher atomic volumes, the logarithmic–linear model continues to exhibit lower MSE relative to both the linear and logarithmic models. Overall, the logarithmic–linear model maintains consistently low error across all atomic volume bins, indicating stable performance over the full range.

These results indicate that while the linear and logarithmic models each capture aspects of the dependence of $S_{vib}$ on atomic volume, neither alone is sufficient to represent the full behavior across the entire range. The logarithmic dependence is qualitatively consistent with expectations from lattice dynamical models, such as the Debye approximation, where $S_{vib}$ is linked to characteristic phonon frequencies that scale with interatomic spacing and volume. However, atomic volume in this dataset acts as an effective descriptor that implicitly reflects multiple underlying factors, including bonding strength, mass, and structural complexity, which are not explicitly accounted for in a simple logarithmic form.

The additional linear term in the logarithmic–linear model captures these systematic deviations from idealized scaling, enabling improved agreement with the data. As a result, the logarithmic–linear model provides a compact yet physically motivated representation that accommodates both the dominant logarithmic trend and secondary contributions arising from variations in material properties. Additional validation through parity analysis (Figure S1) confirms that all models achieve reasonable predictive performance, with the logarithmic–linear model exhibiting the most consistent agreement across the dataset.

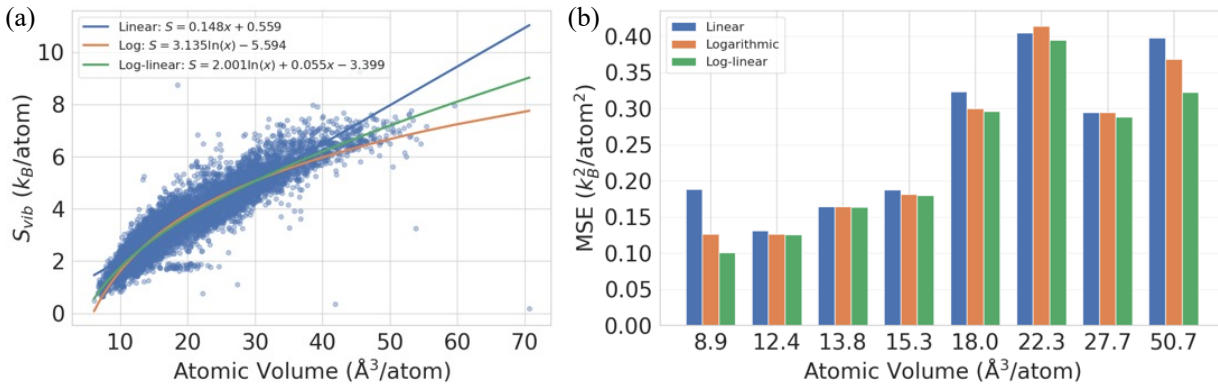

**Figure 3:** (a) Dependence of vibrational entropy ($S_{vib}$) on atomic volume, showing fitted linear, logarithmic, and logarithmic–linear models, with corresponding fitted expressions. (b) Mean squared error (MSE) as a function of atomic volume, evaluated in equal-population bins. The logarithmic–linear model exhibits consistently low error across all regimes, indicating improved stability compared to the linear and logarithmic models. MSE is reported in units of $(k_B/\text{atom})^2$.

### 3.3. Unified temperature-dependent model for vibrational entropy

The dependence of vibrational entropy on atomic volume and temperature was further examined using a temperature-dependent formulation guided by lattice dynamical considerations. In crystalline solids, the temperature dependence of $S_{vib}$ is governed by phonon population statistics.

At low temperatures, the $S_{vib}$ is expected to follow a $T^3$ dependence, consistent with the Debye approximation, reflecting the gradual population of low-frequency acoustic phonon modes. At higher temperatures, as the system approaches the classical limit, $S_{vib}$ exhibits a weaker temperature dependence that can be approximated by a logarithmic form, as suggested by models such as the Einstein approximation.

Motivated by these distinct regimes, a temperature-dependent model was constructed by separating the dataset into low and high-temperature regions using an optimal transition temperature $T_c$ (see Section 2.4). For the present dataset, the optimal transition temperature was found to be $T_c = 200K$. The fitted functional form of the model is given by:

$$S_{vib} = 0.331 \ln(x) + 0.087x + 2.81 \times 10^{-7} T^3 - 1.630 \quad (T < T_c) \tag{4}$$
$$S_{vib} = 1.869 \log(x) + 0.060x + 2.462 \ln(T) - 17.115 \quad (T \geq T_c) \tag{5}$$

where $x$ denotes atomic volume and T represent temperature. These expressions combine physically motivated temperature scaling with structural dependence captured through atomic volume.

The predictive performance of the model is shown in Figure 4 through parity analysis for both training and test datasets. A strong agreement between predicted and actual $S_{vib}$ values is observed across the full range, with data points closely distributed around the diagonal. The reported error metrics further confirm that the model captures the combined influence of temperature and atomic volume with good accuracy.

Overall, this temperature-dependent formulation provides a physically informed description of vibrational entropy, where the low-temperature $T^3$ behavior and high-temperature logarithmic scaling are naturally incorporated within a unified framework.

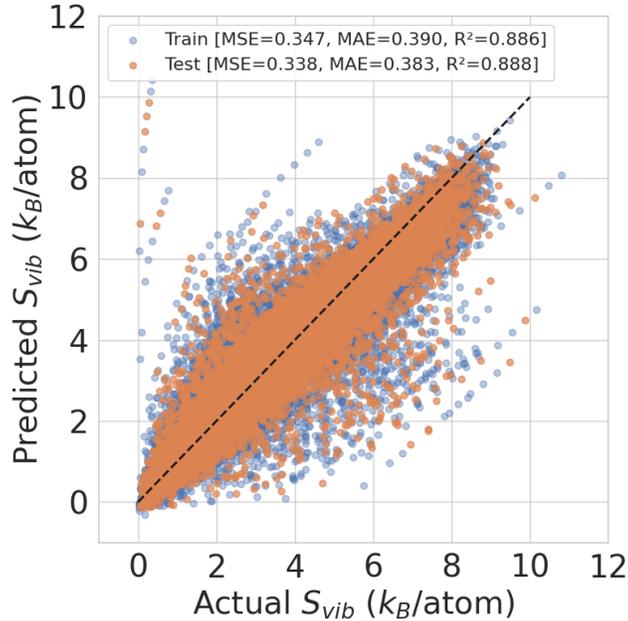

**Figure 4:** Parity plot comparing predicted and actual vibrational entropy ($S_{vib}$) for the temperature-dependent model, evaluated over the temperature range 100–500 K in steps of 50 K. Results are shown for both training and test datasets. The dashed line represents perfect agreement. Mean squared error (MSE), mean absolute error (MAE), and coefficient of determination ($R^2$) are reported for both datasets, with MSE in $(k_B/\text{atom})^2$, MAE in $k_B/\text{atom}$, and $R^2$ dimensionless.

## 4. Conclusion

In this work, we developed a physically informed and computationally efficient framework for predicting vibrational entropy by combining machine learning with simplified analytical modeling. Neural network predictions based on a diverse descriptor set achieve high accuracy, while interpretability analysis reveals atomic volume as the dominant factor governing vibrational entropy. This insight enables the construction of reduced-order models that capture the dependence of vibrational entropy on atomic volume using simple functional forms.

A logarithmic–linear relationship is shown to provide a consistent description across a wide range of materials, reflecting the underlying connection between vibrational entropy and characteristic phonon frequencies. Extending this approach, a temperature-dependent model incorporating $T^3$ behavior at low temperatures and logarithmic scaling at higher temperatures provides a unified description of thermal effects, consistent with established lattice dynamical models.

The resulting framework bridges data-driven modeling and physical interpretation, demonstrating that complex vibrational properties can be described using compact and interpretable relationships. This approach significantly reduces computational cost compared to full phonon calculations and offers a practical route for incorporating vibrational entropy into large-scale materials screening and design.

## Data Availability

The datasets and code used in this study are publicly available at:
https://github.com/shivamt-eng/VibEntropyML

## Acknowledgement

S.T. is grateful to Koushik Pal and Salman Ahmad Khan at IIT Kanpur for useful discussion, The author(s) thank the Anusandhan National Research Foundation (ANRF), India, for financial support through the Prime Minister Early Career Research Grant (PM ECRG) [ANRF/ECRG/2024/000523/ENS]. Computational resources and support provided by PARAM Sanganak under the National Supercomputing Mission, Government of India, at the Indian Institute of Technology Kanpur are gratefully acknowledged.

## Declaration of generative AI and AI-assisted technologies in the writing process

During the preparation of this work, the author(s) used ChatGPT in order to restructure some of the sentences. After using this tool/service, the author(s) reviewed and edited the content as needed and take(s) full responsibility for the content of the publication.